\documentclass[prl,aps,twocolumn,amstex,preprintnumbers,amsmath,amssymb,array,tabularx]{revtex4}
\usepackage{color} % Text coloring
\usepackage{graphicx}% Include figure files
\usepackage{dcolumn}% Align table columns on decimal point
\usepackage{bm}% bold math
\usepackage{lscape} %lanscape/portrait
\usepackage{amsmath}
\usepackage{amsxtra}
\usepackage{tabularx}
\usepackage{footmisc}
\definecolor{AliceBlue}{rgb}{0.94,0.97,1.00}
\definecolor{AntiqueWhite1}{rgb}{1.00,0.94,0.86}
\definecolor{AntiqueWhite2}{rgb}{0.93,0.87,0.80}
\definecolor{AntiqueWhite3}{rgb}{0.80,0.75,0.69}
\definecolor{AntiqueWhite4}{rgb}{0.55,0.51,0.47}
\definecolor{AntiqueWhite}{rgb}{0.98,0.92,0.84}
\definecolor{BlanchedAlmond}{rgb}{1.00,0.92,0.80}
\definecolor{BlueViolet}{rgb}{0.54,0.17,0.89}
\definecolor{CadetBlue1}{rgb}{0.60,0.96,1.00}
\definecolor{CadetBlue2}{rgb}{0.56,0.90,0.93}
\definecolor{CadetBlue3}{rgb}{0.48,0.77,0.80}
\definecolor{CadetBlue4}{rgb}{0.33,0.53,0.55}
\definecolor{CadetBlue}{rgb}{0.37,0.62,0.63}
\definecolor{CornflowerBlue}{rgb}{0.39,0.58,0.93}
\definecolor{DarkBlue}{rgb}{0.00,0.00,0.55}
\definecolor{DarkCyan}{rgb}{0.00,0.55,0.55}
\definecolor{DarkGoldenrod1}{rgb}{1.00,0.73,0.06}
\definecolor{DarkGoldenrod2}{rgb}{0.93,0.68,0.05}
\definecolor{DarkGoldenrod3}{rgb}{0.80,0.58,0.05}
\definecolor{DarkGoldenrod4}{rgb}{0.55,0.40,0.03}
\definecolor{DarkGoldenrod}{rgb}{0.72,0.53,0.04}
\definecolor{DarkGray}{rgb}{0.66,0.66,0.66}
\definecolor{DarkGreen}{rgb}{0.00,0.39,0.00}
\definecolor{DarkGrey}{rgb}{0.66,0.66,0.66}
\definecolor{DarkKhaki}{rgb}{0.74,0.72,0.42}
\definecolor{DarkMagenta}{rgb}{0.55,0.00,0.55}
\definecolor{DarkOliveGreen1}{rgb}{0.79,1.00,0.44}
\definecolor{DarkOliveGreen2}{rgb}{0.74,0.93,0.41}
\definecolor{DarkOliveGreen3}{rgb}{0.64,0.80,0.35}
\definecolor{DarkOliveGreen4}{rgb}{0.43,0.55,0.24}
\definecolor{DarkOliveGreen}{rgb}{0.33,0.42,0.18}
\definecolor{DarkOrange1}{rgb}{1.00,0.50,0.00}
\definecolor{DarkOrange2}{rgb}{0.93,0.46,0.00}
\definecolor{DarkOrange3}{rgb}{0.80,0.40,0.00}
\definecolor{DarkOrange4}{rgb}{0.55,0.27,0.00}
\definecolor{DarkOrange}{rgb}{1.00,0.55,0.00}
\definecolor{DarkOrchid1}{rgb}{0.75,0.24,1.00}
\definecolor{DarkOrchid2}{rgb}{0.70,0.23,0.93}
\definecolor{DarkOrchid3}{rgb}{0.60,0.20,0.80}
\definecolor{DarkOrchid4}{rgb}{0.41,0.13,0.55}
\definecolor{DarkOrchid}{rgb}{0.60,0.20,0.80}
\definecolor{DarkRed}{rgb}{0.55,0.00,0.00}
\definecolor{DarkSalmon}{rgb}{0.91,0.59,0.48}
\definecolor{DarkSeaGreen1}{rgb}{0.76,1.00,0.76}
\definecolor{DarkSeaGreen2}{rgb}{0.71,0.93,0.71}
\definecolor{DarkSeaGreen3}{rgb}{0.61,0.80,0.61}
\definecolor{DarkSeaGreen4}{rgb}{0.41,0.55,0.41}
\definecolor{DarkSeaGreen}{rgb}{0.56,0.74,0.56}
\definecolor{DarkSlateBlue}{rgb}{0.28,0.24,0.55}
\definecolor{DarkSlateGray1}{rgb}{0.59,1.00,1.00}
\definecolor{DarkSlateGray2}{rgb}{0.55,0.93,0.93}
\definecolor{DarkSlateGray3}{rgb}{0.47,0.80,0.80}
\definecolor{DarkSlateGray4}{rgb}{0.32,0.55,0.55}
\definecolor{DarkSlateGray}{rgb}{0.18,0.31,0.31}
\definecolor{DarkSlateGrey}{rgb}{0.18,0.31,0.31}
\definecolor{DarkTurquoise}{rgb}{0.00,0.81,0.82}
\definecolor{DarkViolet}{rgb}{0.58,0.00,0.83}
\definecolor{DeepPink1}{rgb}{1.00,0.08,0.58}
\definecolor{DeepPink2}{rgb}{0.93,0.07,0.54}
\definecolor{DeepPink3}{rgb}{0.80,0.06,0.46}
\definecolor{DeepPink4}{rgb}{0.55,0.04,0.31}
\definecolor{DeepPink}{rgb}{1.00,0.08,0.58}
\definecolor{DeepSkyBlue1}{rgb}{0.00,0.75,1.00}
\definecolor{DeepSkyBlue2}{rgb}{0.00,0.70,0.93}
\definecolor{DeepSkyBlue3}{rgb}{0.00,0.60,0.80}
\definecolor{DeepSkyBlue4}{rgb}{0.00,0.41,0.55}
\definecolor{DeepSkyBlue}{rgb}{0.00,0.75,1.00}
\definecolor{DimGray}{rgb}{0.41,0.41,0.41}
\definecolor{DimGrey}{rgb}{0.41,0.41,0.41}
\definecolor{DodgerBlue1}{rgb}{0.12,0.56,1.00}
\definecolor{DodgerBlue2}{rgb}{0.11,0.53,0.93}
\definecolor{DodgerBlue3}{rgb}{0.09,0.45,0.80}
\definecolor{DodgerBlue4}{rgb}{0.06,0.31,0.55}
\definecolor{DodgerBlue}{rgb}{0.12,0.56,1.00}
\definecolor{FloralWhite}{rgb}{1.00,0.98,0.94}
\definecolor{ForestGreen}{rgb}{0.13,0.55,0.13}
\definecolor{GhostWhite}{rgb}{0.97,0.97,1.00}
\definecolor{GreenYellow}{rgb}{0.68,1.00,0.18}
\definecolor{HotPink1}{rgb}{1.00,0.43,0.71}
\definecolor{HotPink2}{rgb}{0.93,0.42,0.65}
\definecolor{HotPink3}{rgb}{0.80,0.38,0.56}
\definecolor{HotPink4}{rgb}{0.55,0.23,0.38}
\definecolor{HotPink}{rgb}{1.00,0.41,0.71}
\definecolor{IndianRed1}{rgb}{1.00,0.42,0.42}
\definecolor{IndianRed2}{rgb}{0.93,0.39,0.39}
\definecolor{IndianRed3}{rgb}{0.80,0.33,0.33}
\definecolor{IndianRed4}{rgb}{0.55,0.23,0.23}
\definecolor{IndianRed}{rgb}{0.80,0.36,0.36}
\definecolor{LavenderBlush1}{rgb}{1.00,0.94,0.96}
\definecolor{LavenderBlush2}{rgb}{0.93,0.88,0.90}
\definecolor{LavenderBlush3}{rgb}{0.80,0.76,0.77}
\definecolor{LavenderBlush4}{rgb}{0.55,0.51,0.53}
\definecolor{LavenderBlush}{rgb}{1.00,0.94,0.96}
\definecolor{LawnGreen}{rgb}{0.49,0.99,0.00}
\definecolor{LemonChiffon1}{rgb}{1.00,0.98,0.80}
\definecolor{LemonChiffon2}{rgb}{0.93,0.91,0.75}
\definecolor{LemonChiffon3}{rgb}{0.80,0.79,0.65}
\definecolor{LemonChiffon4}{rgb}{0.55,0.54,0.44}
\definecolor{LemonChiffon}{rgb}{1.00,0.98,0.80}
\definecolor{LightBlue1}{rgb}{0.75,0.94,1.00}
\definecolor{LightBlue2}{rgb}{0.70,0.87,0.93}
\definecolor{LightBlue3}{rgb}{0.60,0.75,0.80}
\definecolor{LightBlue4}{rgb}{0.41,0.51,0.55}
\definecolor{LightBlue}{rgb}{0.68,0.85,0.90}
\definecolor{LightCoral}{rgb}{0.94,0.50,0.50}
\definecolor{LightCyan1}{rgb}{0.88,1.00,1.00}
\definecolor{LightCyan2}{rgb}{0.82,0.93,0.93}
\definecolor{LightCyan3}{rgb}{0.71,0.80,0.80}
\definecolor{LightCyan4}{rgb}{0.48,0.55,0.55}
\definecolor{LightCyan}{rgb}{0.88,1.00,1.00}
\definecolor{LightGoldenrod1}{rgb}{1.00,0.93,0.55}
\definecolor{LightGoldenrod2}{rgb}{0.93,0.86,0.51}
\definecolor{LightGoldenrod3}{rgb}{0.80,0.75,0.44}
\definecolor{LightGoldenrod4}{rgb}{0.55,0.51,0.30}
\definecolor{LightGoldenrodYellow}{rgb}{0.98,0.98,0.82}
\definecolor{LightGoldenrod}{rgb}{0.93,0.87,0.51}
\definecolor{LightGray}{rgb}{0.83,0.83,0.83}
\definecolor{LightGreen}{rgb}{0.56,0.93,0.56}
\definecolor{LightGrey}{rgb}{0.83,0.83,0.83}
\definecolor{LightPink1}{rgb}{1.00,0.68,0.73}
\definecolor{LightPink2}{rgb}{0.93,0.64,0.68}
\definecolor{LightPink3}{rgb}{0.80,0.55,0.58}
\definecolor{LightPink4}{rgb}{0.55,0.37,0.40}
\definecolor{LightPink}{rgb}{1.00,0.71,0.76}
\definecolor{LightSalmon1}{rgb}{1.00,0.63,0.48}
\definecolor{LightSalmon2}{rgb}{0.93,0.58,0.45}
\definecolor{LightSalmon3}{rgb}{0.80,0.51,0.38}
\definecolor{LightSalmon4}{rgb}{0.55,0.34,0.26}
\definecolor{LightSalmon}{rgb}{1.00,0.63,0.48}
\definecolor{LightSeaGreen}{rgb}{0.13,0.70,0.67}
\definecolor{LightSkyBlue1}{rgb}{0.69,0.89,1.00}
\definecolor{LightSkyBlue2}{rgb}{0.64,0.83,0.93}
\definecolor{LightSkyBlue3}{rgb}{0.55,0.71,0.80}
\definecolor{LightSkyBlue4}{rgb}{0.38,0.48,0.55}
\definecolor{LightSkyBlue}{rgb}{0.53,0.81,0.98}
\definecolor{LightSlateBlue}{rgb}{0.52,0.44,1.00}
\definecolor{LightSlateGray}{rgb}{0.47,0.53,0.60}
\definecolor{LightSlateGrey}{rgb}{0.47,0.53,0.60}
\definecolor{LightSteelBlue1}{rgb}{0.79,0.88,1.00}
\definecolor{LightSteelBlue2}{rgb}{0.74,0.82,0.93}
\definecolor{LightSteelBlue3}{rgb}{0.64,0.71,0.80}
\definecolor{LightSteelBlue4}{rgb}{0.43,0.48,0.55}
\definecolor{LightSteelBlue}{rgb}{0.69,0.77,0.87}
\definecolor{LightYellow1}{rgb}{1.00,1.00,0.88}
\definecolor{LightYellow2}{rgb}{0.93,0.93,0.82}
\definecolor{LightYellow3}{rgb}{0.80,0.80,0.71}
\definecolor{LightYellow4}{rgb}{0.55,0.55,0.48}
\definecolor{LightYellow}{rgb}{1.00,1.00,0.88}
\definecolor{LimeGreen}{rgb}{0.20,0.80,0.20}
\definecolor{MediumAquamarine}{rgb}{0.40,0.80,0.67}
\definecolor{MediumBlue}{rgb}{0.00,0.00,0.80}
\definecolor{MediumOrchid1}{rgb}{0.88,0.40,1.00}
\definecolor{MediumOrchid2}{rgb}{0.82,0.37,0.93}
\definecolor{MediumOrchid3}{rgb}{0.71,0.32,0.80}
\definecolor{MediumOrchid4}{rgb}{0.48,0.22,0.55}
\definecolor{MediumOrchid}{rgb}{0.73,0.33,0.83}
\definecolor{MediumPurple1}{rgb}{0.67,0.51,1.00}
\definecolor{MediumPurple2}{rgb}{0.62,0.47,0.93}
\definecolor{MediumPurple3}{rgb}{0.54,0.41,0.80}
\definecolor{MediumPurple4}{rgb}{0.36,0.28,0.55}
\definecolor{MediumPurple}{rgb}{0.58,0.44,0.86}
\definecolor{MediumSeaGreen}{rgb}{0.24,0.70,0.44}
\definecolor{MediumSlateBlue}{rgb}{0.48,0.41,0.93}
\definecolor{MediumSpringGreen}{rgb}{0.00,0.98,0.60}
\definecolor{MediumTurquoise}{rgb}{0.28,0.82,0.80}
\definecolor{MediumVioletRed}{rgb}{0.78,0.08,0.52}
\definecolor{MidnightBlue}{rgb}{0.10,0.10,0.44}
\definecolor{MintCream}{rgb}{0.96,1.00,0.98}
\definecolor{MistyRose1}{rgb}{1.00,0.89,0.88}
\definecolor{MistyRose2}{rgb}{0.93,0.84,0.82}
\definecolor{MistyRose3}{rgb}{0.80,0.72,0.71}
\definecolor{MistyRose4}{rgb}{0.55,0.49,0.48}
\definecolor{MistyRose}{rgb}{1.00,0.89,0.88}
\definecolor{NavajoWhite1}{rgb}{1.00,0.87,0.68}
\definecolor{NavajoWhite2}{rgb}{0.93,0.81,0.63}
\definecolor{NavajoWhite3}{rgb}{0.80,0.70,0.55}
\definecolor{NavajoWhite4}{rgb}{0.55,0.47,0.37}
\definecolor{NavajoWhite}{rgb}{1.00,0.87,0.68}
\definecolor{NavyBlue}{rgb}{0.00,0.00,0.50}
\definecolor{OldLace}{rgb}{0.99,0.96,0.90}
\definecolor{OliveDrab1}{rgb}{0.75,1.00,0.24}
\definecolor{OliveDrab2}{rgb}{0.70,0.93,0.23}
\definecolor{OliveDrab3}{rgb}{0.60,0.80,0.20}
\definecolor{OliveDrab4}{rgb}{0.41,0.55,0.13}
\definecolor{OliveDrab}{rgb}{0.42,0.56,0.14}
\definecolor{OrangeRed1}{rgb}{1.00,0.27,0.00}
\definecolor{OrangeRed2}{rgb}{0.93,0.25,0.00}
\definecolor{OrangeRed3}{rgb}{0.80,0.22,0.00}
\definecolor{OrangeRed4}{rgb}{0.55,0.15,0.00}
\definecolor{OrangeRed}{rgb}{1.00,0.27,0.00}
\definecolor{PaleGoldenrod}{rgb}{0.93,0.91,0.67}
\definecolor{PaleGreen1}{rgb}{0.60,1.00,0.60}
\definecolor{PaleGreen2}{rgb}{0.56,0.93,0.56}
\definecolor{PaleGreen3}{rgb}{0.49,0.80,0.49}
\definecolor{PaleGreen4}{rgb}{0.33,0.55,0.33}
\definecolor{PaleGreen}{rgb}{0.60,0.98,0.60}
\definecolor{PaleTurquoise1}{rgb}{0.73,1.00,1.00}
\definecolor{PaleTurquoise2}{rgb}{0.68,0.93,0.93}
\definecolor{PaleTurquoise3}{rgb}{0.59,0.80,0.80}
\definecolor{PaleTurquoise4}{rgb}{0.40,0.55,0.55}
\definecolor{PaleTurquoise}{rgb}{0.69,0.93,0.93}
\definecolor{PaleVioletRed1}{rgb}{1.00,0.51,0.67}
\definecolor{PaleVioletRed2}{rgb}{0.93,0.47,0.62}
\definecolor{PaleVioletRed3}{rgb}{0.80,0.41,0.54}
\definecolor{PaleVioletRed4}{rgb}{0.55,0.28,0.36}
\definecolor{PaleVioletRed}{rgb}{0.86,0.44,0.58}
\definecolor{PapayaWhip}{rgb}{1.00,0.94,0.84}
\definecolor{PeachPuff1}{rgb}{1.00,0.85,0.73}
\definecolor{PeachPuff2}{rgb}{0.93,0.80,0.68}
\definecolor{PeachPuff3}{rgb}{0.80,0.69,0.58}
\definecolor{PeachPuff4}{rgb}{0.55,0.47,0.40}
\definecolor{PeachPuff}{rgb}{1.00,0.85,0.73}
\definecolor{PowderBlue}{rgb}{0.69,0.88,0.90}
\definecolor{RosyBrown1}{rgb}{1.00,0.76,0.76}
\definecolor{RosyBrown2}{rgb}{0.93,0.71,0.71}
\definecolor{RosyBrown3}{rgb}{0.80,0.61,0.61}
\definecolor{RosyBrown4}{rgb}{0.55,0.41,0.41}
\definecolor{RosyBrown}{rgb}{0.74,0.56,0.56}
\definecolor{RoyalBlue1}{rgb}{0.28,0.46,1.00}
\definecolor{RoyalBlue2}{rgb}{0.26,0.43,0.93}
\definecolor{RoyalBlue3}{rgb}{0.23,0.37,0.80}
\definecolor{RoyalBlue4}{rgb}{0.15,0.25,0.55}
\definecolor{RoyalBlue}{rgb}{0.25,0.41,0.88}
\definecolor{SaddleBrown}{rgb}{0.55,0.27,0.07}
\definecolor{SandyBrown}{rgb}{0.96,0.64,0.38}
\definecolor{SeaGreen1}{rgb}{0.33,1.00,0.62}
\definecolor{SeaGreen2}{rgb}{0.31,0.93,0.58}
\definecolor{SeaGreen3}{rgb}{0.26,0.80,0.50}
\definecolor{SeaGreen4}{rgb}{0.18,0.55,0.34}
\definecolor{SeaGreen}{rgb}{0.18,0.55,0.34}
\definecolor{SkyBlue1}{rgb}{0.53,0.81,1.00}
\definecolor{SkyBlue2}{rgb}{0.49,0.75,0.93}
\definecolor{SkyBlue3}{rgb}{0.42,0.65,0.80}
\definecolor{SkyBlue4}{rgb}{0.29,0.44,0.55}
\definecolor{SkyBlue}{rgb}{0.53,0.81,0.92}
\definecolor{SlateBlue1}{rgb}{0.51,0.44,1.00}
\definecolor{SlateBlue2}{rgb}{0.48,0.40,0.93}
\definecolor{SlateBlue3}{rgb}{0.41,0.35,0.80}
\definecolor{SlateBlue4}{rgb}{0.28,0.24,0.55}
\definecolor{SlateBlue}{rgb}{0.42,0.35,0.80}
\definecolor{SlateGray1}{rgb}{0.78,0.89,1.00}
\definecolor{SlateGray2}{rgb}{0.73,0.83,0.93}
\definecolor{SlateGray3}{rgb}{0.62,0.71,0.80}
\definecolor{SlateGray4}{rgb}{0.42,0.48,0.55}
\definecolor{SlateGray}{rgb}{0.44,0.50,0.56}
\definecolor{SlateGrey}{rgb}{0.44,0.50,0.56}
\definecolor{SpringGreen1}{rgb}{0.00,1.00,0.50}
\definecolor{SpringGreen2}{rgb}{0.00,0.93,0.46}
\definecolor{SpringGreen3}{rgb}{0.00,0.80,0.40}
\definecolor{SpringGreen4}{rgb}{0.00,0.55,0.27}
\definecolor{SpringGreen}{rgb}{0.00,1.00,0.50}
\definecolor{SteelBlue1}{rgb}{0.39,0.72,1.00}
\definecolor{SteelBlue2}{rgb}{0.36,0.67,0.93}
\definecolor{SteelBlue3}{rgb}{0.31,0.58,0.80}
\definecolor{SteelBlue4}{rgb}{0.21,0.39,0.55}
\definecolor{SteelBlue}{rgb}{0.27,0.51,0.71}
\definecolor{VioletRed1}{rgb}{1.00,0.24,0.59}
\definecolor{VioletRed2}{rgb}{0.93,0.23,0.55}
\definecolor{VioletRed3}{rgb}{0.80,0.20,0.47}
\definecolor{VioletRed4}{rgb}{0.55,0.13,0.32}
\definecolor{VioletRed}{rgb}{0.82,0.13,0.56}
\definecolor{WhiteSmoke}{rgb}{0.96,0.96,0.96}
\definecolor{YellowGreen}{rgb}{0.60,0.80,0.20}
\definecolor{aliceblue}{rgb}{0.94,0.97,1.00}
\definecolor{antiquewhite}{rgb}{0.98,0.92,0.84}
\definecolor{aquamarine1}{rgb}{0.50,1.00,0.83}
\definecolor{aquamarine2}{rgb}{0.46,0.93,0.78}
\definecolor{aquamarine3}{rgb}{0.40,0.80,0.67}
\definecolor{aquamarine4}{rgb}{0.27,0.55,0.45}
\definecolor{aquamarine}{rgb}{0.50,1.00,0.83}
\definecolor{azure1}{rgb}{0.94,1.00,1.00}
\definecolor{azure2}{rgb}{0.88,0.93,0.93}
\definecolor{azure3}{rgb}{0.76,0.80,0.80}
\definecolor{azure4}{rgb}{0.51,0.55,0.55}
\definecolor{azure}{rgb}{0.94,1.00,1.00}
\definecolor{beige}{rgb}{0.96,0.96,0.86}
\definecolor{bisque1}{rgb}{1.00,0.89,0.77}
\definecolor{bisque2}{rgb}{0.93,0.84,0.72}
\definecolor{bisque3}{rgb}{0.80,0.72,0.62}
\definecolor{bisque4}{rgb}{0.55,0.49,0.42}
\definecolor{bisque}{rgb}{1.00,0.89,0.77}
\definecolor{black}{rgb}{0.00,0.00,0.00}
\definecolor{blanchedalmond}{rgb}{1.00,0.92,0.80}
\definecolor{blue1}{rgb}{0.00,0.00,1.00}
\definecolor{blue2}{rgb}{0.00,0.00,0.93}
\definecolor{blue3}{rgb}{0.00,0.00,0.80}
\definecolor{blue4}{rgb}{0.00,0.00,0.55}
\definecolor{blueviolet}{rgb}{0.54,0.17,0.89}
\definecolor{blue}{rgb}{0.00,0.00,1.00}
\definecolor{brown1}{rgb}{1.00,0.25,0.25}
\definecolor{brown2}{rgb}{0.93,0.23,0.23}
\definecolor{brown3}{rgb}{0.80,0.20,0.20}
\definecolor{brown4}{rgb}{0.55,0.14,0.14}
\definecolor{brown}{rgb}{0.65,0.16,0.16}
\definecolor{burlywood1}{rgb}{1.00,0.83,0.61}
\definecolor{burlywood2}{rgb}{0.93,0.77,0.57}
\definecolor{burlywood3}{rgb}{0.80,0.67,0.49}
\definecolor{burlywood4}{rgb}{0.55,0.45,0.33}
\definecolor{burlywood}{rgb}{0.87,0.72,0.53}
\definecolor{cadetblue}{rgb}{0.37,0.62,0.63}
\definecolor{chartreuse1}{rgb}{0.50,1.00,0.00}
\definecolor{chartreuse2}{rgb}{0.46,0.93,0.00}
\definecolor{chartreuse3}{rgb}{0.40,0.80,0.00}
\definecolor{chartreuse4}{rgb}{0.27,0.55,0.00}
\definecolor{chartreuse}{rgb}{0.50,1.00,0.00}
\definecolor{chocolate1}{rgb}{1.00,0.50,0.14}
\definecolor{chocolate2}{rgb}{0.93,0.46,0.13}
\definecolor{chocolate3}{rgb}{0.80,0.40,0.11}
\definecolor{chocolate4}{rgb}{0.55,0.27,0.07}
\definecolor{chocolate}{rgb}{0.82,0.41,0.12}
\definecolor{coral1}{rgb}{1.00,0.45,0.34}
\definecolor{coral2}{rgb}{0.93,0.42,0.31}
\definecolor{coral3}{rgb}{0.80,0.36,0.27}
\definecolor{coral4}{rgb}{0.55,0.24,0.18}
\definecolor{coral}{rgb}{1.00,0.50,0.31}
\definecolor{cornflowerblue}{rgb}{0.39,0.58,0.93}
\definecolor{cornsilk1}{rgb}{1.00,0.97,0.86}
\definecolor{cornsilk2}{rgb}{0.93,0.91,0.80}
\definecolor{cornsilk3}{rgb}{0.80,0.78,0.69}
\definecolor{cornsilk4}{rgb}{0.55,0.53,0.47}
\definecolor{cornsilk}{rgb}{1.00,0.97,0.86}
\definecolor{cyan1}{rgb}{0.00,1.00,1.00}
\definecolor{cyan2}{rgb}{0.00,0.93,0.93}
\definecolor{cyan3}{rgb}{0.00,0.80,0.80}
\definecolor{cyan4}{rgb}{0.00,0.55,0.55}
\definecolor{cyan}{rgb}{0.00,1.00,1.00}
\definecolor{darkblue}{rgb}{0.00,0.00,0.55}
\definecolor{darkcyan}{rgb}{0.00,0.55,0.55}
\definecolor{darkgoldenrod}{rgb}{0.72,0.53,0.04}
\definecolor{darkgray}{rgb}{0.66,0.66,0.66}
\definecolor{darkgreen}{rgb}{0.00,0.39,0.00}
\definecolor{darkgrey}{rgb}{0.66,0.66,0.66}
\definecolor{darkkhaki}{rgb}{0.74,0.72,0.42}
\definecolor{darkmagenta}{rgb}{0.55,0.00,0.55}
\definecolor{darkolive}{rgb}{0.33,0.42,0.18}
\definecolor{darkorange}{rgb}{1.00,0.55,0.00}
\definecolor{darkorchid}{rgb}{0.60,0.20,0.80}
\definecolor{darkred}{rgb}{0.55,0.00,0.00}
\definecolor{darksalmon}{rgb}{0.91,0.59,0.48}
\definecolor{darksea}{rgb}{0.56,0.74,0.56}
\definecolor{darkslate}{rgb}{0.18,0.31,0.31}
\definecolor{darkslate}{rgb}{0.18,0.31,0.31}
\definecolor{darkslate}{rgb}{0.28,0.24,0.55}
\definecolor{darkturquoise}{rgb}{0.00,0.81,0.82}
\definecolor{darkviolet}{rgb}{0.58,0.00,0.83}
\definecolor{deeppink}{rgb}{1.00,0.08,0.58}
\definecolor{deepsky}{rgb}{0.00,0.75,1.00}
\definecolor{dimgray}{rgb}{0.41,0.41,0.41}
\definecolor{dimgrey}{rgb}{0.41,0.41,0.41}
\definecolor{dodgerblue}{rgb}{0.12,0.56,1.00}
\definecolor{firebrick1}{rgb}{1.00,0.19,0.19}
\definecolor{firebrick2}{rgb}{0.93,0.17,0.17}
\definecolor{firebrick3}{rgb}{0.80,0.15,0.15}
\definecolor{firebrick4}{rgb}{0.55,0.10,0.10}
\definecolor{firebrick}{rgb}{0.70,0.13,0.13}
\definecolor{floralwhite}{rgb}{1.00,0.98,0.94}
\definecolor{forestgreen}{rgb}{0.13,0.55,0.13}
\definecolor{gainsboro}{rgb}{0.86,0.86,0.86}
\definecolor{ghostwhite}{rgb}{0.97,0.97,1.00}
\definecolor{gold1}{rgb}{1.00,0.84,0.00}
\definecolor{gold2}{rgb}{0.93,0.79,0.00}
\definecolor{gold3}{rgb}{0.80,0.68,0.00}
\definecolor{gold4}{rgb}{0.55,0.46,0.00}
\definecolor{goldenrod1}{rgb}{1.00,0.76,0.15}
\definecolor{goldenrod2}{rgb}{0.93,0.71,0.13}
\definecolor{goldenrod3}{rgb}{0.80,0.61,0.11}
\definecolor{goldenrod4}{rgb}{0.55,0.41,0.08}
\definecolor{goldenrod}{rgb}{0.85,0.65,0.13}
\definecolor{gold}{rgb}{1.00,0.84,0.00}
\definecolor{gray0}{rgb}{0.00,0.00,0.00}
\definecolor{gray100}{rgb}{1.00,1.00,1.00}
\definecolor{gray10}{rgb}{0.10,0.10,0.10}
\definecolor{gray11}{rgb}{0.11,0.11,0.11}
\definecolor{gray12}{rgb}{0.12,0.12,0.12}
\definecolor{gray13}{rgb}{0.13,0.13,0.13}
\definecolor{gray14}{rgb}{0.14,0.14,0.14}
\definecolor{gray15}{rgb}{0.15,0.15,0.15}
\definecolor{gray16}{rgb}{0.16,0.16,0.16}
\definecolor{gray17}{rgb}{0.17,0.17,0.17}
\definecolor{gray18}{rgb}{0.18,0.18,0.18}
\definecolor{gray19}{rgb}{0.19,0.19,0.19}
\definecolor{gray1}{rgb}{0.01,0.01,0.01}
\definecolor{gray20}{rgb}{0.20,0.20,0.20}
\definecolor{gray21}{rgb}{0.21,0.21,0.21}
\definecolor{gray22}{rgb}{0.22,0.22,0.22}
\definecolor{gray23}{rgb}{0.23,0.23,0.23}
\definecolor{gray24}{rgb}{0.24,0.24,0.24}
\definecolor{gray25}{rgb}{0.25,0.25,0.25}
\definecolor{gray26}{rgb}{0.26,0.26,0.26}
\definecolor{gray27}{rgb}{0.27,0.27,0.27}
\definecolor{gray28}{rgb}{0.28,0.28,0.28}
\definecolor{gray29}{rgb}{0.29,0.29,0.29}
\definecolor{gray2}{rgb}{0.02,0.02,0.02}
\definecolor{gray30}{rgb}{0.30,0.30,0.30}
\definecolor{gray31}{rgb}{0.31,0.31,0.31}
\definecolor{gray32}{rgb}{0.32,0.32,0.32}
\definecolor{gray33}{rgb}{0.33,0.33,0.33}
\definecolor{gray34}{rgb}{0.34,0.34,0.34}
\definecolor{gray35}{rgb}{0.35,0.35,0.35}
\definecolor{gray36}{rgb}{0.36,0.36,0.36}
\definecolor{gray37}{rgb}{0.37,0.37,0.37}
\definecolor{gray38}{rgb}{0.38,0.38,0.38}
\definecolor{gray39}{rgb}{0.39,0.39,0.39}
\definecolor{gray3}{rgb}{0.03,0.03,0.03}
\definecolor{gray40}{rgb}{0.40,0.40,0.40}
\definecolor{gray41}{rgb}{0.41,0.41,0.41}
\definecolor{gray42}{rgb}{0.42,0.42,0.42}
\definecolor{gray43}{rgb}{0.43,0.43,0.43}
\definecolor{gray44}{rgb}{0.44,0.44,0.44}
\definecolor{gray45}{rgb}{0.45,0.45,0.45}
\definecolor{gray46}{rgb}{0.46,0.46,0.46}
\definecolor{gray47}{rgb}{0.47,0.47,0.47}
\definecolor{gray48}{rgb}{0.48,0.48,0.48}
\definecolor{gray49}{rgb}{0.49,0.49,0.49}
\definecolor{gray4}{rgb}{0.04,0.04,0.04}
\definecolor{gray50}{rgb}{0.50,0.50,0.50}
\definecolor{gray51}{rgb}{0.51,0.51,0.51}
\definecolor{gray52}{rgb}{0.52,0.52,0.52}
\definecolor{gray53}{rgb}{0.53,0.53,0.53}
\definecolor{gray54}{rgb}{0.54,0.54,0.54}
\definecolor{gray55}{rgb}{0.55,0.55,0.55}
\definecolor{gray56}{rgb}{0.56,0.56,0.56}
\definecolor{gray57}{rgb}{0.57,0.57,0.57}
\definecolor{gray58}{rgb}{0.58,0.58,0.58}
\definecolor{gray59}{rgb}{0.59,0.59,0.59}
\definecolor{gray5}{rgb}{0.05,0.05,0.05}
\definecolor{gray60}{rgb}{0.60,0.60,0.60}
\definecolor{gray61}{rgb}{0.61,0.61,0.61}
\definecolor{gray62}{rgb}{0.62,0.62,0.62}
\definecolor{gray63}{rgb}{0.63,0.63,0.63}
\definecolor{gray64}{rgb}{0.64,0.64,0.64}
\definecolor{gray65}{rgb}{0.65,0.65,0.65}
\definecolor{gray66}{rgb}{0.66,0.66,0.66}
\definecolor{gray67}{rgb}{0.67,0.67,0.67}
\definecolor{gray68}{rgb}{0.68,0.68,0.68}
\definecolor{gray69}{rgb}{0.69,0.69,0.69}
\definecolor{gray6}{rgb}{0.06,0.06,0.06}
\definecolor{gray70}{rgb}{0.70,0.70,0.70}
\definecolor{gray71}{rgb}{0.71,0.71,0.71}
\definecolor{gray72}{rgb}{0.72,0.72,0.72}
\definecolor{gray73}{rgb}{0.73,0.73,0.73}
\definecolor{gray74}{rgb}{0.74,0.74,0.74}
\definecolor{gray75}{rgb}{0.75,0.75,0.75}
\definecolor{gray76}{rgb}{0.76,0.76,0.76}
\definecolor{gray77}{rgb}{0.77,0.77,0.77}
\definecolor{gray78}{rgb}{0.78,0.78,0.78}
\definecolor{gray79}{rgb}{0.79,0.79,0.79}
\definecolor{gray7}{rgb}{0.07,0.07,0.07}
\definecolor{gray80}{rgb}{0.80,0.80,0.80}
\definecolor{gray81}{rgb}{0.81,0.81,0.81}
\definecolor{gray82}{rgb}{0.82,0.82,0.82}
\definecolor{gray83}{rgb}{0.83,0.83,0.83}
\definecolor{gray84}{rgb}{0.84,0.84,0.84}
\definecolor{gray85}{rgb}{0.85,0.85,0.85}
\definecolor{gray86}{rgb}{0.86,0.86,0.86}
\definecolor{gray87}{rgb}{0.87,0.87,0.87}
\definecolor{gray88}{rgb}{0.88,0.88,0.88}
\definecolor{gray89}{rgb}{0.89,0.89,0.89}
\definecolor{gray8}{rgb}{0.08,0.08,0.08}
\definecolor{gray90}{rgb}{0.90,0.90,0.90}
\definecolor{gray91}{rgb}{0.91,0.91,0.91}
\definecolor{gray92}{rgb}{0.92,0.92,0.92}
\definecolor{gray93}{rgb}{0.93,0.93,0.93}
\definecolor{gray94}{rgb}{0.94,0.94,0.94}
\definecolor{gray95}{rgb}{0.95,0.95,0.95}
\definecolor{gray96}{rgb}{0.96,0.96,0.96}
\definecolor{gray97}{rgb}{0.97,0.97,0.97}
\definecolor{gray98}{rgb}{0.98,0.98,0.98}
\definecolor{gray99}{rgb}{0.99,0.99,0.99}
\definecolor{gray9}{rgb}{0.09,0.09,0.09}
\definecolor{gray}{rgb}{0.75,0.75,0.75}
\definecolor{green1}{rgb}{0.00,1.00,0.00}
\definecolor{green2}{rgb}{0.00,0.93,0.00}
\definecolor{green3}{rgb}{0.00,0.80,0.00}
\definecolor{green4}{rgb}{0.00,0.55,0.00}
\definecolor{greenyellow}{rgb}{0.68,1.00,0.18}
\definecolor{green}{rgb}{0.00,1.00,0.00}
\definecolor{grey0}{rgb}{0.00,0.00,0.00}
\definecolor{grey100}{rgb}{1.00,1.00,1.00}
\definecolor{grey10}{rgb}{0.10,0.10,0.10}
\definecolor{grey11}{rgb}{0.11,0.11,0.11}
\definecolor{grey12}{rgb}{0.12,0.12,0.12}
\definecolor{grey13}{rgb}{0.13,0.13,0.13}
\definecolor{grey14}{rgb}{0.14,0.14,0.14}
\definecolor{grey15}{rgb}{0.15,0.15,0.15}
\definecolor{grey16}{rgb}{0.16,0.16,0.16}
\definecolor{grey17}{rgb}{0.17,0.17,0.17}
\definecolor{grey18}{rgb}{0.18,0.18,0.18}
\definecolor{grey19}{rgb}{0.19,0.19,0.19}
\definecolor{grey1}{rgb}{0.01,0.01,0.01}
\definecolor{grey20}{rgb}{0.20,0.20,0.20}
\definecolor{grey21}{rgb}{0.21,0.21,0.21}
\definecolor{grey22}{rgb}{0.22,0.22,0.22}
\definecolor{grey23}{rgb}{0.23,0.23,0.23}
\definecolor{grey24}{rgb}{0.24,0.24,0.24}
\definecolor{grey25}{rgb}{0.25,0.25,0.25}
\definecolor{grey26}{rgb}{0.26,0.26,0.26}
\definecolor{grey27}{rgb}{0.27,0.27,0.27}
\definecolor{grey28}{rgb}{0.28,0.28,0.28}
\definecolor{grey29}{rgb}{0.29,0.29,0.29}
\definecolor{grey2}{rgb}{0.02,0.02,0.02}
\definecolor{grey30}{rgb}{0.30,0.30,0.30}
\definecolor{grey31}{rgb}{0.31,0.31,0.31}
\definecolor{grey32}{rgb}{0.32,0.32,0.32}
\definecolor{grey33}{rgb}{0.33,0.33,0.33}
\definecolor{grey34}{rgb}{0.34,0.34,0.34}
\definecolor{grey35}{rgb}{0.35,0.35,0.35}
\definecolor{grey36}{rgb}{0.36,0.36,0.36}
\definecolor{grey37}{rgb}{0.37,0.37,0.37}
\definecolor{grey38}{rgb}{0.38,0.38,0.38}
\definecolor{grey39}{rgb}{0.39,0.39,0.39}
\definecolor{grey3}{rgb}{0.03,0.03,0.03}
\definecolor{grey40}{rgb}{0.40,0.40,0.40}
\definecolor{grey41}{rgb}{0.41,0.41,0.41}
\definecolor{grey42}{rgb}{0.42,0.42,0.42}
\definecolor{grey43}{rgb}{0.43,0.43,0.43}
\definecolor{grey44}{rgb}{0.44,0.44,0.44}
\definecolor{grey45}{rgb}{0.45,0.45,0.45}
\definecolor{grey46}{rgb}{0.46,0.46,0.46}
\definecolor{grey47}{rgb}{0.47,0.47,0.47}
\definecolor{grey48}{rgb}{0.48,0.48,0.48}
\definecolor{grey49}{rgb}{0.49,0.49,0.49}
\definecolor{grey4}{rgb}{0.04,0.04,0.04}
\definecolor{grey50}{rgb}{0.50,0.50,0.50}
\definecolor{grey51}{rgb}{0.51,0.51,0.51}
\definecolor{grey52}{rgb}{0.52,0.52,0.52}
\definecolor{grey53}{rgb}{0.53,0.53,0.53}
\definecolor{grey54}{rgb}{0.54,0.54,0.54}
\definecolor{grey55}{rgb}{0.55,0.55,0.55}
\definecolor{grey56}{rgb}{0.56,0.56,0.56}
\definecolor{grey57}{rgb}{0.57,0.57,0.57}
\definecolor{grey58}{rgb}{0.58,0.58,0.58}
\definecolor{grey59}{rgb}{0.59,0.59,0.59}
\definecolor{grey5}{rgb}{0.05,0.05,0.05}
\definecolor{grey60}{rgb}{0.60,0.60,0.60}
\definecolor{grey61}{rgb}{0.61,0.61,0.61}
\definecolor{grey62}{rgb}{0.62,0.62,0.62}
\definecolor{grey63}{rgb}{0.63,0.63,0.63}
\definecolor{grey64}{rgb}{0.64,0.64,0.64}
\definecolor{grey65}{rgb}{0.65,0.65,0.65}
\definecolor{grey66}{rgb}{0.66,0.66,0.66}
\definecolor{grey67}{rgb}{0.67,0.67,0.67}
\definecolor{grey68}{rgb}{0.68,0.68,0.68}
\definecolor{grey69}{rgb}{0.69,0.69,0.69}
\definecolor{grey6}{rgb}{0.06,0.06,0.06}
\definecolor{grey70}{rgb}{0.70,0.70,0.70}
\definecolor{grey71}{rgb}{0.71,0.71,0.71}
\definecolor{grey72}{rgb}{0.72,0.72,0.72}
\definecolor{grey73}{rgb}{0.73,0.73,0.73}
\definecolor{grey74}{rgb}{0.74,0.74,0.74}
\definecolor{grey75}{rgb}{0.75,0.75,0.75}
\definecolor{grey76}{rgb}{0.76,0.76,0.76}
\definecolor{grey77}{rgb}{0.77,0.77,0.77}
\definecolor{grey78}{rgb}{0.78,0.78,0.78}
\definecolor{grey79}{rgb}{0.79,0.79,0.79}
\definecolor{grey7}{rgb}{0.07,0.07,0.07}
\definecolor{grey80}{rgb}{0.80,0.80,0.80}
\definecolor{grey81}{rgb}{0.81,0.81,0.81}
\definecolor{grey82}{rgb}{0.82,0.82,0.82}
\definecolor{grey83}{rgb}{0.83,0.83,0.83}
\definecolor{grey84}{rgb}{0.84,0.84,0.84}
\definecolor{grey85}{rgb}{0.85,0.85,0.85}
\definecolor{grey86}{rgb}{0.86,0.86,0.86}
\definecolor{grey87}{rgb}{0.87,0.87,0.87}
\definecolor{grey88}{rgb}{0.88,0.88,0.88}
\definecolor{grey89}{rgb}{0.89,0.89,0.89}
\definecolor{grey8}{rgb}{0.08,0.08,0.08}
\definecolor{grey90}{rgb}{0.90,0.90,0.90}
\definecolor{grey91}{rgb}{0.91,0.91,0.91}
\definecolor{grey92}{rgb}{0.92,0.92,0.92}
\definecolor{grey93}{rgb}{0.93,0.93,0.93}
\definecolor{grey94}{rgb}{0.94,0.94,0.94}
\definecolor{grey95}{rgb}{0.95,0.95,0.95}
\definecolor{grey96}{rgb}{0.96,0.96,0.96}
\definecolor{grey97}{rgb}{0.97,0.97,0.97}
\definecolor{grey98}{rgb}{0.98,0.98,0.98}
\definecolor{grey99}{rgb}{0.99,0.99,0.99}
\definecolor{grey9}{rgb}{0.09,0.09,0.09}
\definecolor{grey}{rgb}{0.75,0.75,0.75}
\definecolor{honeydew1}{rgb}{0.94,1.00,0.94}
\definecolor{honeydew2}{rgb}{0.88,0.93,0.88}
\definecolor{honeydew3}{rgb}{0.76,0.80,0.76}
\definecolor{honeydew4}{rgb}{0.51,0.55,0.51}
\definecolor{honeydew}{rgb}{0.94,1.00,0.94}
\definecolor{hotpink}{rgb}{1.00,0.41,0.71}
\definecolor{indianred}{rgb}{0.80,0.36,0.36}
\definecolor{ivory1}{rgb}{1.00,1.00,0.94}
\definecolor{ivory2}{rgb}{0.93,0.93,0.88}
\definecolor{ivory3}{rgb}{0.80,0.80,0.76}
\definecolor{ivory4}{rgb}{0.55,0.55,0.51}
\definecolor{ivory}{rgb}{1.00,1.00,0.94}
\definecolor{khaki1}{rgb}{1.00,0.96,0.56}
\definecolor{khaki2}{rgb}{0.93,0.90,0.52}
\definecolor{khaki3}{rgb}{0.80,0.78,0.45}
\definecolor{khaki4}{rgb}{0.55,0.53,0.31}
\definecolor{khaki}{rgb}{0.94,0.90,0.55}
\definecolor{lavenderblush}{rgb}{1.00,0.94,0.96}
\definecolor{lavender}{rgb}{0.90,0.90,0.98}
\definecolor{lawngreen}{rgb}{0.49,0.99,0.00}
\definecolor{lemonchiffon}{rgb}{1.00,0.98,0.80}
\definecolor{lightblue}{rgb}{0.68,0.85,0.90}
\definecolor{lightcoral}{rgb}{0.94,0.50,0.50}
\definecolor{lightcyan}{rgb}{0.88,1.00,1.00}
\definecolor{lightgoldenrod}{rgb}{0.93,0.87,0.51}
\definecolor{lightgoldenrod}{rgb}{0.98,0.98,0.82}
\definecolor{lightgray}{rgb}{0.83,0.83,0.83}
\definecolor{lightgreen}{rgb}{0.56,0.93,0.56}
\definecolor{lightgrey}{rgb}{0.83,0.83,0.83}
\definecolor{lightpink}{rgb}{1.00,0.71,0.76}
\definecolor{lightsalmon}{rgb}{1.00,0.63,0.48}
\definecolor{lightsea}{rgb}{0.13,0.70,0.67}
\definecolor{lightsky}{rgb}{0.53,0.81,0.98}
\definecolor{lightslate}{rgb}{0.47,0.53,0.60}
\definecolor{lightslate}{rgb}{0.47,0.53,0.60}
\definecolor{lightslate}{rgb}{0.52,0.44,1.00}
\definecolor{lightsteel}{rgb}{0.69,0.77,0.87}
\definecolor{lightyellow}{rgb}{1.00,1.00,0.88}
\definecolor{limegreen}{rgb}{0.20,0.80,0.20}
\definecolor{linen}{rgb}{0.98,0.94,0.90}
\definecolor{magenta1}{rgb}{1.00,0.00,1.00}
\definecolor{magenta2}{rgb}{0.93,0.00,0.93}
\definecolor{magenta3}{rgb}{0.80,0.00,0.80}
\definecolor{magenta4}{rgb}{0.55,0.00,0.55}
\definecolor{magenta}{rgb}{1.00,0.00,1.00}
\definecolor{maroon1}{rgb}{1.00,0.20,0.70}
\definecolor{maroon2}{rgb}{0.93,0.19,0.65}
\definecolor{maroon3}{rgb}{0.80,0.16,0.56}
\definecolor{maroon4}{rgb}{0.55,0.11,0.38}
\definecolor{maroon}{rgb}{0.69,0.19,0.38}
\definecolor{mediumaquamarine}{rgb}{0.40,0.80,0.67}
\definecolor{mediumblue}{rgb}{0.00,0.00,0.80}
\definecolor{mediumorchid}{rgb}{0.73,0.33,0.83}
\definecolor{mediumpurple}{rgb}{0.58,0.44,0.86}
\definecolor{mediumsea}{rgb}{0.24,0.70,0.44}
\definecolor{mediumslate}{rgb}{0.48,0.41,0.93}
\definecolor{mediumspring}{rgb}{0.00,0.98,0.60}
\definecolor{mediumturquoise}{rgb}{0.28,0.82,0.80}
\definecolor{mediumviolet}{rgb}{0.78,0.08,0.52}
\definecolor{midnightblue}{rgb}{0.10,0.10,0.44}
\definecolor{mintcream}{rgb}{0.96,1.00,0.98}
\definecolor{mistyrose}{rgb}{1.00,0.89,0.88}
\definecolor{moccasin}{rgb}{1.00,0.89,0.71}
\definecolor{navajowhite}{rgb}{1.00,0.87,0.68}
\definecolor{navyblue}{rgb}{0.00,0.00,0.50}
\definecolor{navy}{rgb}{0.00,0.00,0.50}
\definecolor{oldlace}{rgb}{0.99,0.96,0.90}
\definecolor{olivedrab}{rgb}{0.42,0.56,0.14}
\definecolor{orange1}{rgb}{1.00,0.65,0.00}
\definecolor{orange2}{rgb}{0.93,0.60,0.00}
\definecolor{orange3}{rgb}{0.80,0.52,0.00}
\definecolor{orange4}{rgb}{0.55,0.35,0.00}
\definecolor{orangered}{rgb}{1.00,0.27,0.00}
\definecolor{orange}{rgb}{1.00,0.65,0.00}
\definecolor{orchid1}{rgb}{1.00,0.51,0.98}
\definecolor{orchid2}{rgb}{0.93,0.48,0.91}
\definecolor{orchid3}{rgb}{0.80,0.41,0.79}
\definecolor{orchid4}{rgb}{0.55,0.28,0.54}
\definecolor{orchid}{rgb}{0.85,0.44,0.84}
\definecolor{palegoldenrod}{rgb}{0.93,0.91,0.67}
\definecolor{palegreen}{rgb}{0.60,0.98,0.60}
\definecolor{paleturquoise}{rgb}{0.69,0.93,0.93}
\definecolor{paleviolet}{rgb}{0.86,0.44,0.58}
\definecolor{papayawhip}{rgb}{1.00,0.94,0.84}
\definecolor{peachpuff}{rgb}{1.00,0.85,0.73}
\definecolor{peru}{rgb}{0.80,0.52,0.25}
\definecolor{pink1}{rgb}{1.00,0.71,0.77}
\definecolor{pink2}{rgb}{0.93,0.66,0.72}
\definecolor{pink3}{rgb}{0.80,0.57,0.62}
\definecolor{pink4}{rgb}{0.55,0.39,0.42}
\definecolor{pink}{rgb}{1.00,0.75,0.80}
\definecolor{plum1}{rgb}{1.00,0.73,1.00}
\definecolor{plum2}{rgb}{0.93,0.68,0.93}
\definecolor{plum3}{rgb}{0.80,0.59,0.80}
\definecolor{plum4}{rgb}{0.55,0.40,0.55}
\definecolor{plum}{rgb}{0.87,0.63,0.87}
\definecolor{powderblue}{rgb}{0.69,0.88,0.90}
\definecolor{purple1}{rgb}{0.61,0.19,1.00}
\definecolor{purple2}{rgb}{0.57,0.17,0.93}
\definecolor{purple3}{rgb}{0.49,0.15,0.80}
\definecolor{purple4}{rgb}{0.33,0.10,0.55}
\definecolor{purple}{rgb}{0.63,0.13,0.94}
\definecolor{red1}{rgb}{1.00,0.00,0.00}
\definecolor{red2}{rgb}{0.93,0.00,0.00}
\definecolor{red3}{rgb}{0.80,0.00,0.00}
\definecolor{red4}{rgb}{0.55,0.00,0.00}
\definecolor{red}{rgb}{1.00,0.00,0.00}
\definecolor{rosybrown}{rgb}{0.74,0.56,0.56}
\definecolor{royalblue}{rgb}{0.25,0.41,0.88}
\definecolor{saddlebrown}{rgb}{0.55,0.27,0.07}
\definecolor{salmon1}{rgb}{1.00,0.55,0.41}
\definecolor{salmon2}{rgb}{0.93,0.51,0.38}
\definecolor{salmon3}{rgb}{0.80,0.44,0.33}
\definecolor{salmon4}{rgb}{0.55,0.30,0.22}
\definecolor{salmon}{rgb}{0.98,0.50,0.45}
\definecolor{sandybrown}{rgb}{0.96,0.64,0.38}
\definecolor{seagreen}{rgb}{0.18,0.55,0.34}
\definecolor{seashell1}{rgb}{1.00,0.96,0.93}
\definecolor{seashell2}{rgb}{0.93,0.90,0.87}
\definecolor{seashell3}{rgb}{0.80,0.77,0.75}
\definecolor{seashell4}{rgb}{0.55,0.53,0.51}
\definecolor{seashell}{rgb}{1.00,0.96,0.93}
\definecolor{sienna1}{rgb}{1.00,0.51,0.28}
\definecolor{sienna2}{rgb}{0.93,0.47,0.26}
\definecolor{sienna3}{rgb}{0.80,0.41,0.22}
\definecolor{sienna4}{rgb}{0.55,0.28,0.15}
\definecolor{sienna}{rgb}{0.63,0.32,0.18}
\definecolor{skyblue}{rgb}{0.53,0.81,0.92}
\definecolor{slateblue}{rgb}{0.42,0.35,0.80}
\definecolor{slategray}{rgb}{0.44,0.50,0.56}
\definecolor{slategrey}{rgb}{0.44,0.50,0.56}
\definecolor{snow1}{rgb}{1.00,0.98,0.98}
\definecolor{snow2}{rgb}{0.93,0.91,0.91}
\definecolor{snow3}{rgb}{0.80,0.79,0.79}
\definecolor{snow4}{rgb}{0.55,0.54,0.54}
\definecolor{snow}{rgb}{1.00,0.98,0.98}
\definecolor{springgreen}{rgb}{0.00,1.00,0.50}
\definecolor{steelblue}{rgb}{0.27,0.51,0.71}
\definecolor{tan1}{rgb}{1.00,0.65,0.31}
\definecolor{tan2}{rgb}{0.93,0.60,0.29}
\definecolor{tan3}{rgb}{0.80,0.52,0.25}
\definecolor{tan4}{rgb}{0.55,0.35,0.17}
\definecolor{tan}{rgb}{0.82,0.71,0.55}
\definecolor{thistle1}{rgb}{1.00,0.88,1.00}
\definecolor{thistle2}{rgb}{0.93,0.82,0.93}
\definecolor{thistle3}{rgb}{0.80,0.71,0.80}
\definecolor{thistle4}{rgb}{0.55,0.48,0.55}
\definecolor{thistle}{rgb}{0.85,0.75,0.85}
\definecolor{tomato1}{rgb}{1.00,0.39,0.28}
\definecolor{tomato2}{rgb}{0.93,0.36,0.26}
\definecolor{tomato3}{rgb}{0.80,0.31,0.22}
\definecolor{tomato4}{rgb}{0.55,0.21,0.15}
\definecolor{tomato}{rgb}{1.00,0.39,0.28}
\definecolor{turquoise1}{rgb}{0.00,0.96,1.00}
\definecolor{turquoise2}{rgb}{0.00,0.90,0.93}
\definecolor{turquoise3}{rgb}{0.00,0.77,0.80}
\definecolor{turquoise4}{rgb}{0.00,0.53,0.55}
\definecolor{turquoise}{rgb}{0.25,0.88,0.82}
\definecolor{violetred}{rgb}{0.82,0.13,0.56}
\definecolor{violet}{rgb}{0.93,0.51,0.93}
\definecolor{wheat1}{rgb}{1.00,0.91,0.73}
\definecolor{wheat2}{rgb}{0.93,0.85,0.68}
\definecolor{wheat3}{rgb}{0.80,0.73,0.59}
\definecolor{wheat4}{rgb}{0.55,0.49,0.40}
\definecolor{wheat}{rgb}{0.96,0.87,0.70}
\definecolor{whitesmoke}{rgb}{0.96,0.96,0.96}
\definecolor{white}{rgb}{1.00,1.00,1.00}
\definecolor{yellow1}{rgb}{1.00,1.00,0.00}
\definecolor{yellow2}{rgb}{0.93,0.93,0.00}
\definecolor{yellow3}{rgb}{0.80,0.80,0.00}
\definecolor{yellow4}{rgb}{0.55,0.55,0.00}
\definecolor{yellowgreen}{rgb}{0.60,0.80,0.20}
\definecolor{yellow}{rgb}{1.00,1.00,0.00}

\DeclareGraphicsExtensions{.jpg,.pdf,.eps,.gif}

\begin{document}

\title{Quantifying the Stacking Registry Matching in Layered Materials}

\author{$\mbox{Oded Hod}$}
\affiliation{School of Chemistry, The Sackler Faculty of Exact
Sciences, Tel Aviv University, Tel Aviv 69978, Israel}

\date{\today}

\begin{abstract}
A detailed account of a recently developed method [Marom {\it et al.,
Phys. Rev. Lett.} {\bf 105}, 046801 (2010)] to quantify the registry
mismatch in layered materials is presented. The registry index, which
was originally defined for planar hexagonal boron-nitride, is extended
to treat graphitic systems and generalized to describe multi-layered
nanotubes. It is shown that using simple geometric considerations it
is possible to capture the complex physical features of interlayer
sliding in layered materials. The intuitive nature of the presented
model and the efficiency of the related computations suggest that the
method can be used as a powerful characterization tool for interlayer
interactions in complex layered systems.
\end{abstract}

\maketitle
 
%\newpage 
%%%%%%%%%%%%%%%%%%%%%%%%%%%%%%%%%%%%%%%%%%%%%%%%%%%%%%%%%%%%%%%%%%%%%%%%%%%
%%%%%%%                       BODY OF TEXT                          %%%%%%%
%%%%%%%%%%%%%%%%%%%%%%%%%%%%%%%%%%%%%%%%%%%%%%%%%%%%%%%%%%%%%%%%%%%%%%%%%%%

%%%%%%%%%%%%%%%%%%%%%%%%%%%%%%%%%%%%%%%%%%%%%%%%%%%%%%%%%%%%%%%%%%%%%%%%%%%

\renewcommand{\thefootnote}{\fnsymbol{footnote}}

\section{Introduction}
Single layers of atomically thin molecular structures have attracted
vast attention in the past few years. Most commonly, they appear in
quasi-one dimensional~\cite{Iijima1991, Tenne1992, Rubio1994,
Chopra1995, Loiseau1996, Tenne2004, Tenne2006, Golberg2007,
saito_book, dress-book} and quasi-two dimensional~\cite{Novoselov2004,
Novoselov2005, Zhang2005, Berger2006, Zhi2009, Li2009, Bunch2007,
Geim2007} forms. The physical properties of these unique structures
depend on their specific geometry and chemical composition being
organic,~\cite{Iijima1991, saito_book, dress-book}
inorganic,~\cite{Tenne1992, Tenne2004, Tenne2006, Rubio1994,
Chopra1995, Loiseau1996, Novoselov2005, Golberg2007, Zhi2009, Li2009}
or mixed.~\cite{Stephan1994, Miyamoto1994_I, Miyamoto1994_II,
Weng-Sieh1995, Redlich1996, Terrones1996, Suenaga1997, Blase1997,
Zhang1997, Yap2009, Ci2010, Rubio2010} When stacked together to form
multi-layered structures, the physical properties of the individual
layers may be considerably altered via interlayer
interactions.~\cite{Tanaka1997, Kwon1998, Palser1999, Paulson2000,
Liu2003, Zhang2004, Kolmogorov2005, Novoselov2006, McCann2006,
Ooi2006, Koskilinna2006, Charlier2007, Graf2007, Nagapriya2008,
Marom2010} Due to the different nature of the intra- and inter-layer
interactions, the resulting layered systems often present highly
anisotropic properties. Within each layer covalent bonding usually
results in relatively high strength and stiffness~\cite{saito_book,
dress-book, Yakobson2001, Golberg2007} along the direction of the
layer and, in some cases, in efficient electronic~\cite{saito_book,
dress-book, Novoselov2004, Morozov2008} and heat~\cite{Berber2000,
Che2000, Tang2006, Balandin2008, Ghosh2008, Morooka2008, Ghosh2009,
Hu2009, Lan2009, Jiang2009, Ouyang2009, Xu2009, Ouyang2010}
transport. In contrast, between the layers long-range dispersion and
electrostatic interactions produce weaker and more flexible mechanical
properties~\cite{Cumings2000, Kolmogorov2000, Zheng2002, Legoas2003,
Fennimore2003, Deshpande2006} and result in reduced transport
capabilities.

\begin{figure}[h]
  \begin{center}
    \epsfxsize=9.00cm \epsffile{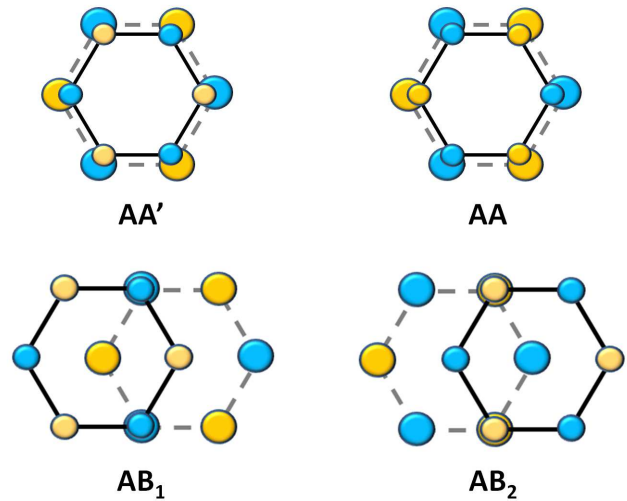}
  \end{center}
  \caption{High symmetry stacking modes of $h$-BN. Upper left panel:
  the optimal $AA'$ stacking mode. Upper right panel: the fully
  eclipsed $AA$ stacking mode. Lower left panel: the meta-stable
  $AB_1$ stacking mode. Lower right panel: the high energy $AB_2$
  stacking mode. Blue(yellow) circles in the on-line version represent
  boron(nitrogen) atoms. Dashed gray lines with large circles
  represents a lower $h$-BN layer and solid black lines with small
  circles represents an upper layer.}
  \label{Fig: Stacking modes}
\end{figure}

An important factor governing the physical properties of multi-layered
materials is the registry matching between the layers. Depending on
the nature of the interlayer interactions, different layered materials
present different optimal stacking modes. As an example, the polar
nature of the interlayer B-N covalent bond in hexagonal boron-nitride
($h$-BN) results in considerable Coulomb interactions between atomic
sites belonging to different layers. These, in turn, dictate an
optimal $AA'$ stacking mode, where a boron(nitrogen) atom in one layer
resides above a nitrogen(boron) atom in its adjacent layers (see
Fig.~\ref{Fig: Stacking modes}). Graphite is iso-electronic to $h$-BN
and has the same hexagonal structure within each layer. It is
therefore tempting to assume that both systems present a similar
optimal stacking mode. Nevertheless, due to the lack of bond
polarization in the homo-nuclear intralayer covalent bonds of graphene
the dominant interlayer interactions in graphite and in few-layered
graphene (FLG) are dispersion forces. These, in turn, lead to an
optimal $AB$ stacking mode, where a carbon atom in one graphene layer
resides on top of a hollow site of the corresponding adjacent
layers. The picture becomes even more complicated when considering
multi-walled nanotubes. In such systems, apart from the specific
chemical composition of the nanotube, which dictates the nature of the
interlayer interactions, curvature differences and different rolling
orientations (chiral angles) of adjacent layers result in complex
registry matching and mismatching patterns often regarded as Moir\'e
fringes.~\cite{Nagapriya2008}

It is therefore clear that registry matching plays an important role
in dictating the electronic properties of layered materials and in the
interlayer sliding physics of such systems.~\cite{Drexler1986,
Drexler1987, Mate1987, Drexler1992, Merkle1993, Falvo1998, Buldum1999,
Schall2000, Yu2000, Liu2003, Rivera2003, Zhang2004, Zou2006,
Marom2010} Nevertheless, previous studies concerning such effects have
regarded the registry matching in qualitative terms labeling it as
``bad'', ``good'', or ``optimal'' according to the relative energetic
stability of the different stacking modes as calculated by either
force fields~\cite{Kolmogorov2000, Zhang2007} or via electronic
structure theories.~\cite{Liu2003, Carlson2007}

Recently, a quantitative measure of the registry matching in planar
$h$-BN was proposed.~\cite{Marom2010} Based on intuitive geometrical
considerations and common knowledge regarding the nature of the
interlayer interactions in $h$-BN, a simple model was derived, which
predicts the relative stability of different stacking modes. It was
shown that the main features of the interlayer sliding physics in this
material can be captured by this simplified model thus shedding light
on the main factors that govern this complex process. It is the
purpose of this paper to give a detailed account of this registry
index model, extending it to other planar layered materials such as
graphite, and generalizing its applicability to multi-walled nanotube
structures.

%In the following section a detailed explanation of the physical
%considerations leading to the definition of the registry index in
%$h$-BN is presented. Next, we extend this definition to the case of
%planar graphitic systems. Finally, the model is generalized to treat
%multi-walled nanotube systems.

\section{Registry index in $h$-BN}
In order to define a quantitative measure of the registry mismatch in
planar $h$-BN it is important to understand the nature of the
interlayer interactions in this system. Three important forces should
be taken into account:

\begin{enumerate}

\item \emph{Dispersion interactions:} Van der Waals-London forces play
a central role in the physics of molecular stacking. While being
weaker than the intralayer covalent bonding, these induced
dipole-dipole interactions are responsible for anchoring the different
layers of a multi-layered material at the appropriate interlayer
distance.~\cite{Marom2010} Nevertheless, it was recently shown that
dispersion forces are relatively insensitive to the specific
interlayer arrangement of different stacking modes in $h$-BN and thus
have little effect on the interlayer sliding
potential.~\cite{Marom2010}
  
\item \emph{Electrostatic interactions between ionic cores:} The
dominant factor governing the interlayer sliding potential of $h$-BN
are electrostatic attractions and repulsions between the partially
charged atomic centers on adjacent layers. Due to the difference in
electronegativity of the two atoms, the boron bears a partial positive
charge whereas the nitrogen has a partial negative
charge.~\cite{Yamamura1997, Liu2003, Marom2010} Based on these
observations and on basic electrostatic considerations, one may deduce
that optimal registry is achieved at the $AA'$ mode which maximizes
the interlayer N-B Coulomb attractions and minimizes the corresponding
B-B and N-N repulsions (see upper left panel of Fig.~\ref{Fig:
Stacking modes}). Similarly, the worst stacking mode is the $AA$ mode,
where the $h$-BN sheets are completely eclipsed and Coulomb repulsions
between atomic centers are maximal (upper right panel of the figure).

\item \emph{Charge densities overlap:} Another factor that may
influence the relative stability of different stacking modes is the
electrostatic and Pauli repulsion due to partial overlap of the
electron densities surrounding the boron and nitrogen atomic
centers.~\cite{Yamamura1997, Liu2003, Barreiro2008, Marom2010} The
electron cloud associated with the boron atom is smaller than that of
the nitrogen atom. one may therefore expect that the $AB_1$ stacking
mode with eclipsed boron atoms (lower left panel of Fig.~\ref{Fig:
Stacking modes}), will be more energetically favorable than the
corresponding $AB_2$ mode with eclipsed nitrogen atoms (lower right
panel of the figure). Interestingly, though the interlayer electron
densities overlap is small~\cite{Yamamura1997} it seems to have a
considerable effect on the relative stability of the $AB_1$ and $AB_2$
stacking modes.~\cite{Liu2003, Marom2010, Schon2010}

\end{enumerate}

\input{epsf}

\begin{figure}[h]
  \begin{center}
    \epsfxsize=6.50cm \epsffile{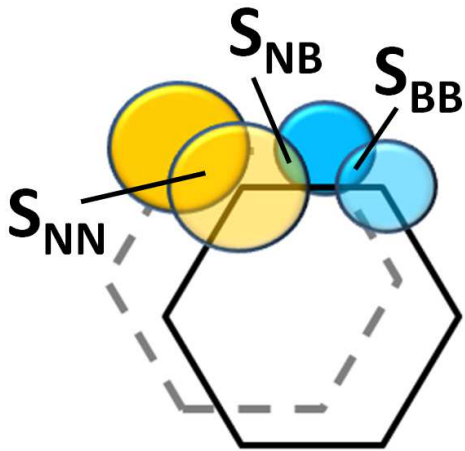}
  \end{center}
  \caption{Registry index definition of the overlap area between
  circles assigned to atomic positions in the upper layer (transparent
  circles) and their lower layer counterparts (opaque circles). The
  circles representing the atomic centers in Fig.~\ref{Fig: Stacking
  modes} were omitted for clarity. The color code is the same as in
  Fig.~\ref{Fig: Stacking modes}}
  \label{Fig: Overlap}
\end{figure}

Having identified the main interactions involved in $h$-BN interlayer
coupling we may now turn to define a quantitative measure of the
registry matching in this system. Similar to the total energy of the
system, we are interested in a simple numerical index which will
obtain a minimum value for the optimal $AA'$ stacking mode and a
maximum value for the worst $AA$ mode. To this end, we ascribe to each
atom in the unit cell a circle centered around its position (see
Fig.~\ref{Fig: Overlap}). Focusing on two adjacent layers, we see that
the projection of a circle assigned to a specific atom located on one
of the layers may overlap with circles associated with atoms of the
same and/or opposite types on the other layer. We mark by $S_{ij}$ the
overlaps between two such circles, one associated with an $i$ atom on
one layer and the other with a $j$ atom on the second layer, $i$ and
$j$ being either B or N. It is now clear that the sum $S_{NN} + S_{BB}
- S_{NB}$ complies with our requirement of obtaining a
minimum(maximum) value at the $AA'$($AA$) stacking mode, where
$S_{NB}$ is maximal(minimal) and $S_{BB}$ and $S_{NN}$ are
minimal(maximal). By normalizing this sum to be limited to the range
[0,1] we obtain the registry index ($RI$) for $h$-BN:
\begin{equation}
  RI_{\mbox{$h$-BN}} = \frac{(S_{NN}-S_{NN}^{AA'}) +
    (S_{BB}-S_{BB}^{AA'}) -
    (S_{NB}-S_{NB}^{AA'})}{(S_{NN}^{AA}-S_{NN}^{AA'}) +
    (S_{BB}^{AA}-S_{BB}^{AA'}) - (S_{NB}^{AA}-S_{NB}^{AA'})}.
  \label{Eq: Registry index}
\end{equation}
Here, $S_{ij}^{AA'}$ and $S_{ij}^{AA}$ are the corresponding overlaps
at the $AA'$ and $AA$ stacking modes, respectively.

\input{epsf}

This definition was obtained based on the knowledge that in $h$-BN
perfect(worst) registry is achieved at the $AA'$($AA$) stacking mode
due to electrostatic interactions between partially charged atomic
centers. As stated above, the effect of interlayer overlap of charge
densities influences the relative stability of the $AB_1$ and $AB_2$
configurations. In order to take this into account we assign different
radii to circles associated with boron ($r_B$) and nitrogen ($r_N$)
atoms. This may be viewed as a simplified representation of the
difference in atomic radii between the partially negatively charged
nitrogen and positively charged boron atoms. For simplicity, we fix
the circle associated with the nitrogen atom to half the B-N bond
length, and use the ratio $\frac{r_B}{r_N}$ as a free parameter. By
choosing $\frac{r_B}{r_N} < 1$, $RI$ obtains a lower value for the
$AB_1$ stacking mode with respect to the $AB_2$ mode thus reproducing
the physical requirement.

It is now possible to plot the $RI$ at different stacking modes and
compare the resulting surface obtained from simple geometric
considerations with the sliding energy surface obtained from advance
electronic structure calculations. Such a comparison was recently
presented,~\cite{Marom2010} showing a remarkable agreement between
density functional theory (DFT) results obtain via the PBE density
functional approximation~\cite{pbe} augmented with the Tkatchenko-
Scheffler Van der Waals correction~\cite{Tkatchenko2009, Marom2010_II}
and the $RI$ model with $\frac{r_B}{r_N}=0.3$. This exemplifies the
fact that the $h$-BN sliding process is governed by registry mismatch
via electrostatic interactions and validates our assumption regarding
the choice of different boron and nitrogen circle radii within the
$RI$ model. We therefore conclude that the $RI$, as defined above, can
be used to characterize the different stacking modes of $h$-BN. The
question arises whether this simple geometric model can be extended to
treat other layered materials such as graphite, and generalized to
more complex structures such as nanotubes. Should the answer to this
question be positive, one would be able to gain valuable physical
intuition regarding such layered materials and characterize their
relative interlayer configurations at a fraction of the computational
cost of current electronic structure methods.

A clue to the answer to this question can be found in related recent
studies using geometric considerations for the description of halogen
atoms and rare gases adsorbed on (111) metal
surfaces.~\cite{Tkatchenko2006, Tkatchenko2006_II} In what follows we
show how the current $RI$ model can be extended and generalized as
suggested above.

\section{Registry index in graphitic materials}
We start by showing that the $RI$ model is not limited to the case of
$h$-BN and can be extended to treat other planar layered materials. We
consider the case of graphite or FLG. As stated above, because of the
homo-nuclear nature of the bonds in these systems no charge
polarization occurs. Therefore, the main factors governing the sliding
physics are dispersion forces and overlap of charge densities. Due to
the lack of electrostatic forces, the $AA$ stacking mode, which is
equivalent to the stable $AA'$ mode in $h$-BN, is found to be a
maximum on the interlayer potential energy surface. Furthermore, the
$AB$ configuration, which minimizes charge densities overlap, is the
optimal stacking mode of graphite. Interestingly, the interlayer
distance, which is mostly influenced by dispersion
interactions,~\cite{Rydberg2003, Akdim2003, Ortmann2006, Marini2006,
Spanu2009, Pakarinen2009, Marom2010, Marom2010_II} is found to be very
similar to that of $h$-BN (3.35~$\AA$ vs. 3.33~$\AA$ in $h$-BN).

Based on these observations one can now extend the registry index
definition to treat graphitic materials. The graphitic $RI$ should
obtain a minimum value for the optimal $AB$ stacking mode of graphene
and a maximum value for the worst $AA$ mode of this system. As for the
case of $h$-BN, we ascribe to each atom in the unit cell a circle of
radius $r_C=0.5d_{cc}$ centered around its position, where $d_{cc}$ is
the intralayer carbon-carbon covalent bond length. The overlap between
two such circles, one associated with a carbon atom on one layer and
the other with a carbon atom on the second layer, is then marked by
$S_{CC}$.  Naturally, if the $RI$ is chosen to be proportional to
$S_{CC}$ it will comply with the requirement of obtaining a
minimum(maximum) value at the $AB$($AA$) stacking mode. By normalizing
$RI$ to be limited to the range [0,1] we obtain the following
definition:

\begin{equation}
  RI_{\mbox{graphitic}} =
  \frac{S_{CC}-S_{CC}^{AB}}{S_{CC}^{AA}-S_{CC}^{AB}}.
  \label{Eq: Registry index graphitic}
\end{equation}
Here, $S_{CC}^{AA}$ and $S_{CC}^{AB}$ are the overlaps at the $AA$ and
$AB$ stacking modes,respectively.

\input{epsf}
\begin{figure}[h]
  \begin{center}
    \epsfxsize=8.5cm \epsffile{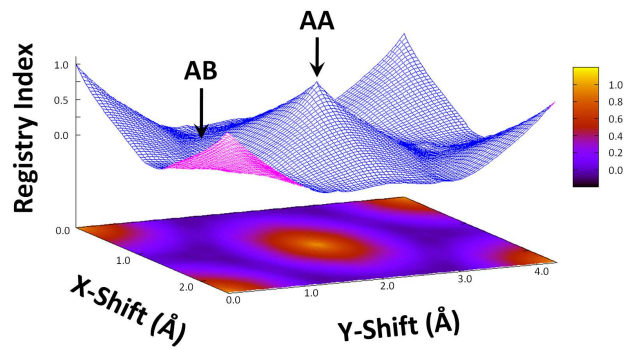}
  \end{center}
  \caption{Registry index surface of double layered graphene.}
  \label{Fig: DL-graphene}
\end{figure}

In Fig.~\ref{Fig: DL-graphene} the registry index surface is presented
as a function of relative interlayer sliding parallel to the basal
planes of a bilayer graphene system. Comparing to recent molecular
dynamics~\cite{Zhang2007} and dispersion-augmented tight-binding
calculations,~\cite{Carlson2007} it is found that the $RI$ landscape
fully captures all the important features of the interlayer sliding
physics of graphene. Furthermore, our results are consistent with
recent experimental and theoretical investigations showing an
orientation dependent sliding resistance in graphitic
systems.~\cite{Dienwiebel2004, Filippov2008} This proves that the $RI$
concept is of general nature and can be readily extended to
characterize the interlayer interactions in a variety of layered
materials. We now turn to describe how this model can be further
generalized to treat more complex structures.

\section{Registry index in multi-walled nanotubes}
Planar layered materials usually have a compact unit cell, which can
be readily treated using standard electronic structure methods within
periodic boundary conditions calculations. On the other hand, despite
their reduced dimensionality, even achiral nanotubes present
relatively large unit cells. This is especially true in the case of
multi-walled nanotubes (MWNTs) where often one finds that, apart from
the smallest bilayer systems, they are beyond the reach of
state-of-the-art electronic structure methods. It is therefore
desirable to generalize the $RI$ defined above to treat tubular
structures. Once such a generalization is established, it can be used
as an efficient and reliable characterization tool for the relative
stability of different inter-tube configurations.

Since nearest neighboring layers interactions are the most important
factors governing the relative arrangement of the different layers
within a MWNT, we generalize the $RI$ to the case of double-walled
nanotubes (DWNT). This allows the investigation of the isolated
layer-layer interactions, which are at the basis of the multi-layered
system behavior.

\input{epsf}
\begin{figure}[h]
  \begin{center}
    \epsfxsize=8.60cm \epsffile{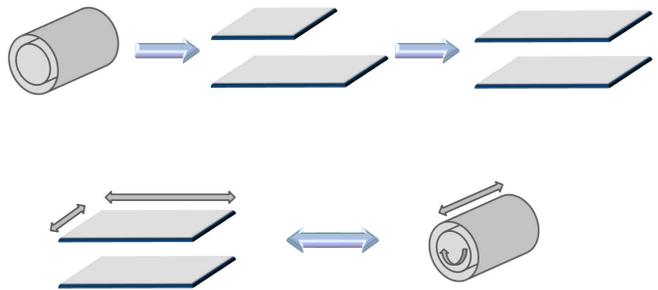}
  \end{center}
  \caption{Schematic representation of the procedure to calculate the
  $RI$ of DWNT systems. First, the two layers are unrolled. Next, the
  narrower sheet (inner layer) is stretched to match the width of the
  wider sheet. Finally, circles are placed around the atomic positions
  and the RI surface is calculated for different relative positions of
  the unrolled layers. These, in turn, are equivalent to relative
  telescoping and rotation of the tubular system.}
  \label{Fig: RI-nanotube}
\end{figure}

The procedure to calculate the $RI$ in DWNTs is schematically
presented in Fig.~\ref{Fig: RI-nanotube}. We start by cutting the two
layers along a given line parallel to the axis of the tube. Next the
layers are unrolled to form planar sheets of different width. Then,
the narrower sheet (unrolled inner tube) is stretched to match the
width of the wider sheet (unrolled outer tube), thus taking into
account the effect of curvature on the registry mismatch between the
two layers. Finally, circles are placed around the atomic centers of
the two layers and the $RI$ is calculated using Eq.~\ref{Eq: Registry
index} (or Eq.~\ref{Eq: Registry index graphitic}) for different
interlayer shifts parallel to the basal planes of the two layers. The
resulting $RI$ surface corresponds to relative telescoping and
rotation of the two tubes within the DWNT.

Similar to the case of $h$-BN, we can now compare the $RI$ surfaces to
the results obtained by DFT calculations.  To this end, we perform a
set of DFT calculations with the \textsc{Gaussian} suite of
programs.~\cite{gdv_short, Gaussian} The local spin density (LSDA),
PBE,~\cite{pbe} and HSE06~\cite{Heyd2003, Heyd2006}
exchange-correlation functional approximations are used together with
the double-$\zeta$ polarized 6-31G** Gaussian basis
set~\cite{Hariharan1973}. As discussed above, Dispersion interactions
play a major role in anchoring the layers of $h$-BN at the appropriate
interlayer distance. Nevertheless, in the case of DWNTs, the
interlayer distance is fixed by the differences of curvature between
the two tubes which are set by the tubes indices. Hence, the effects
of dispersion interactions on the interlayer sliding energy, which
have been shown to be of minor importance in $h$-BN,~\cite{Marom2010}
are neglected in the present work.

Results for three representative double-walled boron-nitride nanotubes
(DWBNNT) are presented: $(5,5)@(10,10)$, $(6,6)@(11,11)$, and
$(6,0)@(14,0)$, where the notation $(n_1,m_1)@(n_2,m_2)$ stands for an
inner $(n_1,m_1)$ tube placed within an outer $(n_2,m_2)$ tube. Unlike
the case of planar $h$-BN, DWNTs present a wide range of possible
structures. The two tubes may differ in chiralities, a factor that may
considerably alter their registry matching and result in
orders-of-magnitude differences in their sliding energy
corrugation. Interestingly, even for achiral tubes of the same kind
(armchair or zigzag) two types of systems can be constructed: the
first (type-I) resulting from rolling two $h$-BN sheets in the $AA'$
stacking mode, and the other (type-II) resulting from rolling two $AA$
stacked $h$-BN layers. It should be noted that one such achiral DWNT
may be obtained from its counterpart by switching the identities of
the boron and nitrogen atoms in one of the layers. As shown below,
once the chiralities and types of the two tubes are set, changing
diameters of the tubes, even while fixing the inter-tube distance, has
remarkable impact on the registry matching between the layers and
their sliding energy surface corrugation.

\input{epsf}
\begin{figure}[h]
  \begin{center}
    \epsfxsize=8.75cm \epsffile{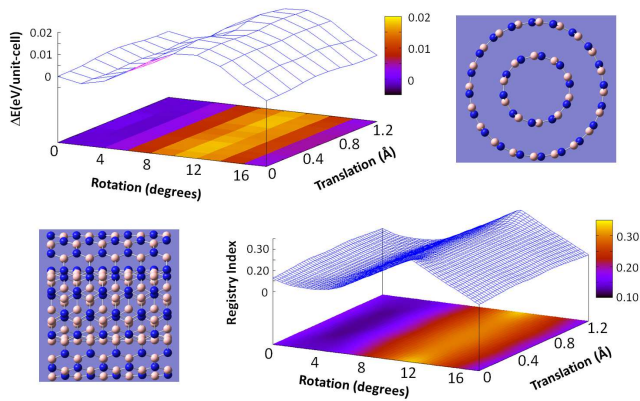}
  \end{center}
  \caption{Rotation-telescoping energy landscape of the type-I
  (5,5)@(10,10) armchair DWBNNT. Upper left panel: Relative total
  energies of different inter-tube configurations calculated using DFT
  at the PBE/6-31G** level of theory. Lower right panel: Registry
  index surface calculated using the procedure described in the
  text. Axial and side views of the system are shown to emphasize the
  effects of curvature on the registry mismatch between the two
  tubes.}
  \label{Fig: (5,5)@(10,10)}
\end{figure}

In Fig.~\ref{Fig: (5,5)@(10,10)} the results for the type-I
$(5,5)@(10,10)$ system are presented. The tubes are formed by rolling
two $AA'$ stacked layers, while fixing the B-N distance to be
$\sim$1.44~$\AA$. No geometry optimizations are performed. The
resulting distance between the tubes is $\sim$3.44~$\AA$ which is
similar to the equilibrium interlayer distance of $h$-BN of
3.33~$\AA$.~\cite{Solozhenko1995} In the upper left panel of the
figure results obtained at the PBE/6-31G** level of theory are
presented. Similar results have been obtained using the LSDA and HSE06
functionals (not shown). The interlayer potential energy is found to
be much more sensitive to relative rotations of the two armchair tubes
than to telescoping parallel to the tubes axis. The corrugation
energy, which is defined as the maximal amplitude of energy changes
between different interlayer relative positions, is found to be $\sim
0.02$~eV/unit-cell. The corresponding $RI$ surface presented in the
lower right panel of the figure reproduces all of these effects while
capturing even the fine details of the sliding energy surface
landscape.

\input{epsf}
\begin{figure}[h]
  \begin{center}
    \epsfxsize=8.75cm \epsffile{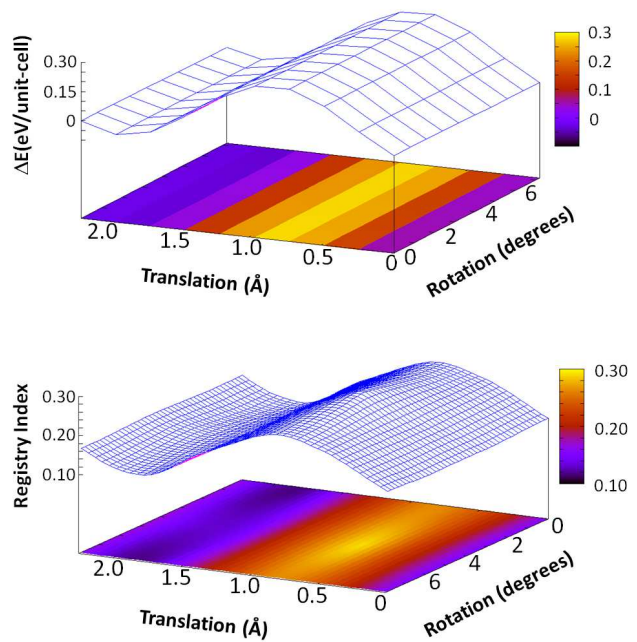}
  \end{center}
  \caption{Rotation-telescoping energy landscape of the type-I
  (6,0)@(14,0) zigzag DWBNNT. Upper panel: Relative total
  energies of different inter-tube configurations calculated using DFT
  at the LSDA/6-31G** level of theory. Lower panel: Registry index
  surface calculated using the procedure described in the text.}
  \label{Fig: (6,0)@(14,0)}
\end{figure}

Fig.~\ref{Fig: (6,0)@(14,0)} presents similar results, obtained at the
LSDA/6-31G** level of theory, for the type-I $(6,0)@(14,0)$ zigzag DWNT
system. As can be seen from the upper left panel, the sliding energy
surface is very similar to that obtained for the (5,5)@(10,10)
armchair system. Two important differences are apparent: (i) the role
of the axes is interchanged (ii) the corrugation energy for the zigzag
system is found to be an order of magnitude larger than that of the
armchair system. The reason for the latter difference is the smaller
interlayer distance of 3.18~$\AA$ in the zigzag DWNT
system. Nevertheless, as in the case of the armchair DWNT, the $RI$
landscape fully captures all the details of the sliding energy surface
obtained via DFT calculations.

To exemplify the complexity of the sliding physics of DWBNNT
systems,~\cite{Merkle1993} the $(6,6)@(11,11)$ is considered as
well. This armchair DWNT has the same interlayer distance as the
$(5,5)@(10,10)$ system considered above. One may naively expect that
the sliding energy landscape of the two systems, which have the same
chirality, type, and interlayer distance, would be similar. As can be
seen in the left panel of Fig.~\ref{Fig: (6,6)@(11,11)}, this is not
the case. The Corrugation energy of the $(6,6)@(11,11)$ is found to be
an order of magnitude smaller than that of its $(5,5)@(10,10)$
counterpart. In fact, the energy differences between relative tube
positions are calculated to be smaller than $0.002$~eV/unti-cell,
which is beyond the accuracy of our DFT calculations. Accordingly, the
agreement between the $RI$ landscape (right panel of the figure) and
the DFT results is not as good as those obtained for the
$(5,5)@(10,10)$ and $(6,0)@(14,0)$ systems. Consistent with the
reduction in the corrugation energy, the magnitude of the $RI$
variations is reduces by more than an order of magnitude as well. It
should be stated that the $RI$ remains a valid quantity to describe
the registry matching in this system. Furthermore, the amplitude of
the $RI$ variations may serve as an indication to the ability of DFT
calculations to accurately describe the interlayer sliding
landscape. It remains to be shown whether in such cases of extremely
small corrugation energy the sliding physics is still dominated by the
registry mismatch. An answer to this question can be given only with
more accurate electronic structure calculations including the detailed
effects of dispersion interactions, which may have an important
contribution in these situations.~\cite{Shtogun2010, Tkatchenko2007}

\input{epsf}
\begin{figure}[h]
  \begin{center}
    \epsfxsize=8.75cm \epsffile{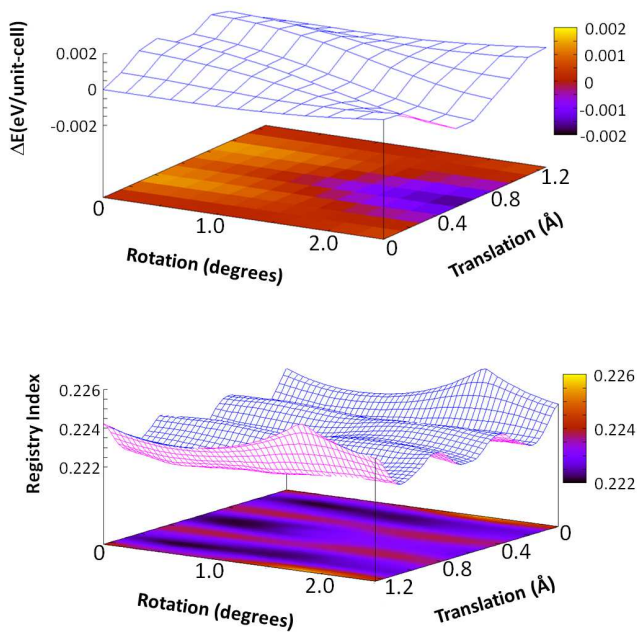}
  \end{center}
  \caption{Rotation-telescoping energy landscape of the type-I
  (6,6)@(11,11) DWBNNT. Upper panel: Relative total energies of
  different inter-tube configurations calculated using DFT at the
  LSDA/6-31G** level of theory. Lower panel: Registry index surface
  calculated using the procedure described in the text.}
  \label{Fig: (6,6)@(11,11)}
\end{figure}

\begin{figure}[h]
  \begin{center}
    \epsfxsize=8.75cm \epsffile{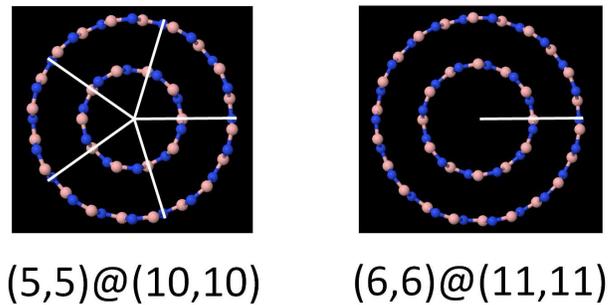}
  \end{center}
  \caption{Comparison of the recurrence frequency of on-top B-N
  stacking arrangements in type-I (5,5)@(10,10) and (6,6)@(11,11)
  DWBNNTs.}
  \label{Fig: Recurrence frequency}
\end{figure}

In order to better understand these differences in the corrugation
energy and $RI$ variations between the (5,5)@(10,10) and (6,6)@(11,11)
DWBNNT we take a closer look at their symmetry characteristics. We
choose an inter-tube arrangement which has a perfect on-top stacking
between a given boron atom on one wall and a nitride atom on the other
wall (see white lines in Fig.~\ref{Fig: Recurrence frequency}). It is
now possible to define a recurrence frequency as the number of times
such an on-top stacking appears along the tube circumference at that
given geometry.  Since a $(n,n)$ boron-nitride nanotube has a $n$-fold
rotational symmetry around the axis of the tube, the recurrence
frequency of a $(n_1,n_1)@(n_2,n_2)$ system is given by
$gcd(n_1,n_2)$, where $gcd$ stands for the greatest common
divisor. For the two armchair DWNTs considered above, the recurrence
frequencies are $gcd(5,10)=5$ and $gcd(6,11)=1$ for the
$(5,5)@(10,10)$ and $(6,6)@(11,11)$ systems, respectively. Naturally,
as the recurrence frequency grows, the $RI$ (and the total energy) of
the on-top configuration decreases, and the corrugation of the sliding
energy surface increases, explaining why the $(5,5)@(10,10)$ system
presents a considerably higher corrugation energy than the
$(6,6)@(11,11)$ DWBNNT.

\input{epsf}

Similar considerations can be used to characterize DWNT of other
chiralities showing the sensitivity of the sliding energy surface of a
DWBNNT to the specific identity of its layers. Nevertheless, the
extension of the $RI$ model to treat tubular structures proves to be a
reliable tool for the quantification of the registry matching between
the layers. Therefore, it can be used to characterize the interlayer
potential and identify optimal interlayer configurations of very large
multi-walled nanotubes that are beyond the reach of current DFT
calculations.

\section{Summary}
A new methodology to quantify the registry matching in layered
materials, based on simple geometric considerations, was
presented. The registry index, which was originally developed to
describe the stacking registry in planar $h$-BN, was extended to treat
graphitic materials and generalized to describe multi-walled
nanotubes. Even in the challenging case of double-walled boron-nitride
nanotubes, the $RI$ model was able to capture the important physical
features of the interlayer sliding up to fine details. This marks the
method as a powerful characterization tool for interlayer interactions
in complex layered systems while giving intuitive insights regarding
the nature of the interlayer couplings.

%***************************************************************

%{\Large \bf  Acknowledgments}

\section{Acknowledgments}
The author would like to thank Prof. Leeor Kronik and Prof. Ernesto
Joselevich from the Weizmann Institute of Science, Dr. Alexandre
Tkatchenko from the Fritz-Haber Institute, Dr. Noa Marom from the
University of Texas at Austin, and Prof. Michael Urbakh from Tel-Aviv
University for many intriguing discussions on the subject. This work
was supported by the Israel Science Foundation under grant N$^o$
1313/08 , and the Center for Nanoscience and Nanotechnology at
Tel-Aviv University. The research leading to these results has
received funding from the European Community's Seventh Framework
Programme FP7/2007-2013 under grant agreement N$^o$ 249225.

%***************************************************************

\bibliographystyle{prsty} \bibliography{Registry}
\end{document}